\documentclass[page-classic]{epl2} 
\usepackage{bbold}

\title{Scarring in open chaotic systems: The local density of states}

\author{D. Lippolis} 

\institute{
       Institute for Applied Systems Analysis, Jiangsu University, Zhenjiang 212013, China             
}

\abstract{
Chaotic Hamiltonians are known to follow Random Matrix Theory (RMT) ensembles in the apparent randomness of their spectra and wavefunction statistics. Deviations from RMT also do occur, however, due to system-specific properties, or as quantum signatures of classical chaos. Scarring, for instance, is the enhancement of 
wavefunction intensity near classical periodic orbits, and it can be characterized by a local density of states (or local spectrum) that clearly deviates from RMT expectations, by   
exhibiting a peaked envelope, which has been described semiclassically.    
Here, the system is connected to an opening,
the local density of states is introduced for the resulting non-Hermitian chaotic Hamiltonian, and estimated a priori in terms of the Green's function of the closed system and the open channels.
The predictions obtained are tested on  quantum maps coupled both to a single-channel opening
 and to a Fresnel-type continuous opening.
The main outcome is that strong coupling to the opening gradually suppresses the energy dependence of the local density of states due to scarring, and restores RMT behavior.   
}

\begin{document}

\maketitle

\section{Introduction}


 Spectral- and wavefunction statistics play a prominent role in the detection, understanding, and classification of quantum chaotic behavior. It is by now established knowledge that, depending on symmetries and thus on conserved quantities, quantized chaotic Hamiltonians modelling complex systems may be described using random matrices~\cite{Mehta,stoeck,Weiden86}.
Analogously, eigenfunctions may be regarded as superpositions of waves with fixed energy (wavevector) and random amplitudes and phases~\cite{Berry77}.

On the other hand,
localization arises in quantum chaos as an interesting anomaly from the expected random-wave 
behavior, and it is often ascribed to classical invariants~\cite{Haake,ErCaSa09,Ketz18}.  
An example that will be central in the present work is quantum scarring, that is enhancement or suppression of wavefunction intensity near
unstable periodic orbits~\cite{Hel84}, and consequent systematic deviation of the spectral statistics from Random Matrix Theory (RMT).
Scars have been shown to significantly affect the wavefunction statistics~\cite{Bogo88,Borondo94}, as well as the local spectrum, and some short- and long-time dynamical properties~\cite{KapHel98,Kap98}. Quantitative understanding of the semiclassical dynamics behind scarring has also been used to construct efficient basis functions for the diagonalization of the full quantum Hamiltonians and propagators~\cite{Verg00,VergCar01,Cr_Lee,Lee_Cr,Nonne03,Verg12,Borondo17}.
Over the years, the investigation of scars has further developed into experimental observations for example at microwave frequencies~\cite{stoeck}, as well as generalizations to pseudointegrable systems~\cite{Bogo04,Bogo06,Dietz07}, and effects on trace formulae~\cite{Dietz08,Dietz10,Dietz12}.
More recently, `strong' quantum scars have been detected and described in integrable systems with impurities~\cite{StrongScars16,ContScars17}.     

Open systems generate a wide class of problems that has greatly concerned the quantum chaos community in the last two decades~\cite{KuhlRev05,APTrev13,Novaes}. Scattering is of fundamental importance to understand for example 
transport properties in mesoscopic systems~\cite{Jalabert16,Huang15}, ionization or photoabsorption in complex atoms and molecules~\cite{FyodAl98,Hern17}, the effects of absorption  in devices such as microwave cavities~\cite{MS03,Kuhl05,Gradoni15} or microwave networks~\cite{Dietz16,Dietz17,Stock18},  and directed radiation from dielectric optical microcavities~\cite{WierHent,Susumu,CaoWier,XueFeng} with an underlying chaotic classical dynamics. It is then natural 
to investigate dynamical properties and wavefunction statistics of open chaotic systems 
to address both fundamental and practical questions, such as understanding deviations from the RMT results due to absorption~\cite{PniShap,Sebaetal,FyodSav04,FyodSav05,FyodSavRev05,Bartetal05,SavLgMrts06,
FyodSav12,FyodSav15,Savin17,Savin18}, or how scars are modified by the opening~\cite{keat_supp,WisCar08,LRLK,Prado18}. 
Regarding the latter issue,
scarring is often assessed by visual inspection of real- and phase-space (e.g. Husimi or Wigner) distributions, or by 
counting statistics of the wavefunction intensities.  
The purpose of the present paper is instead to  
study the local density of states~\cite{KapHel98,Kap98,EfetPrig93,Beenakker94}, that is an observable that yields information on the whole spectrum, and
is directly related to the autocorrelation function by a simple Fourier transform. 

More precisely, this letter aims at: $i)$ estimating the local density of states of a non-Hermitian Hamiltonian \textit{a priori} from the Green's function of the closed system and the open channels, and hence, for a fully-developed chaotic system, to extend the semiclassical predictions of Kaplan and Heller's scar theory~\cite{Kap98,KapHel98} to open systems, in non-perturbative regime; $ii)$ studying how absorption affects the local density of states and therefore the dynamics through the predictions obtained, validated by numerical simulations of quantum maps connected to a single-channel opening, and to a Fresnel-type continuous
opening, typical of optical microcavities.
As main result, when the opening is strongly coupled to the probe state centered on the periodic orbit, absorption tends to suppress the spectral signature of a scar: the
characteristic peaked, energy-dependent local density of states flattens toward a uniform distribution, so that RMT behavior is gradually restored.



\section{The local density of states}
Let us first
write the local density of 
states 
in terms of the Green's function:
\begin{equation}
S(E,x) =  -\mathrm{Im} \langle x|G(E)|x\rangle 
\, .
  \label{rewldos}
  \end{equation} 
As said, the focus is on chaotic Hamiltonians coupled with a number of open channels,
of the type~\cite{ISSO94} 
\begin{equation}
H = H_0 -i\Gamma\sum_c|a_c\rangle\langle a_c| ,
\label{multcH}
\end{equation}
where $\Gamma$ is a dimensionless parameter that controls the coupling with the continuum. 
Assume there are $k$ open channels. For convenience, the following notation is adopted~\cite{Shankar}:
let 
\begin{equation}
 A = \left(|a_1\rangle, |a_2\rangle, ..., |a_k\rangle\right), \hspace{1cm} 
 A^\dagger = \left( \begin{array}{r}
\langle a_1| \\ \langle a_2| \\ ... \\ \langle a_k|
\end{array} \right), 
 \label{Matofkets}
 \end{equation}
 be matrices of kets, each representing an open channel. 
We may now use a self-consistent equation to express the Green's function in terms of the resolvent
of the closed system:
\begin{equation}
G(E) = G_0(E) - i\Gamma G(E)\sum_c|a_c\rangle\langle a_c|G_0(E)  = 
 G_0(E) - i\Gamma G(E)AA^\dagger G_0(E)
 \,.
 \label{selfconst}
 \end{equation}
Then the Green's function can be formally expanded in $G_0(E)$ according to the 
above recursion relation,
to obtain~\cite{SokZel89,SokZel92}
\begin{eqnarray}
 \nonumber
G(E) &=& 
G_0(E) - i\Gamma G_0(E)A\left[ \mathbb{1}  - i\Gamma A^\dagger G_0(E)A - 
\Gamma^2 A^\dagger G_0(E)AA^\dagger G_0(E)A +... \right]A^\dagger G_0(E)
\\ 
&=& G_0(E) -  i\Gamma G_0(E)A \left[\mathbb{1} + i\Gamma A^\dagger G_0(E)A\right]^{-1}
A^\dagger G_0(E) .
\label{Dyson}
\end{eqnarray}
In this notation, $A^\dagger G_0A$ is a $k\times k$ matrix of entries
$\left[A^\dagger G_0(E)A\right]_{ij}= \langle a_i|G_0(E)|a_j\rangle$. 
At this point,  
the local density of states may be written as
\begin{equation}
S(E,x) =-\mathrm{Im}\left[
\langle x|G_0(E)|x\rangle - 
 i\Gamma\langle x|G_0(E)A \left(\mathbb{1} + i\Gamma A^\dagger G_0(E)A\right)^{-1}
A^\dagger G_0(E)|x\rangle\right]
\, .
\label{Dysldos}
\end{equation}
\section{Single-channel opening}
Equation~(\ref{Dysldos}) meets our first goal of writing the local density of states (or, equivalently, the local Green's function) of the open system in terms of its closed analog and the open channels.  
In order to gain intuition on the meaning and implications of this result,  
the analysis is now restricted to a single-channel opening: the non-Hermitian Hamiltonian~(\ref{multcH}) becomes  
$H = H_0 -i\Gamma |a\rangle\langle a|$ ,
and the previous derivation in this case yields the local density of states
\begin{equation}
S(E,x) = -\mathrm{Im} 
\left[\langle x|G_0(E)|x\rangle - 
 i\Gamma\frac{|\langle x|G_0(E)|a\rangle|^2}{1+i\Gamma\langle a|G_0(E)|a\rangle}
\right] 
 \, .
\label{Dysldos1c}
 \end{equation}
As in the multiple-channel result, the dependence of Eq.~(\ref{Dysldos1c}) on the local density of states of the closed system (the first term) and on the Green's function of the open channel $\left(\langle a|G_0(E)|a\rangle\right)$ is apparent. In addition to that, though, the single-channel form reveals the `interaction' term $\langle x|G_0(E)|a\rangle$, which connects the probe state $|x\rangle$ to the open channel $|a\rangle$ through the quantum propagator. In order for the local density of states to be affected at all by the opening, that amplitude must be non-negligible. 

Let us consider the special case where $|a\rangle=|x\rangle$, meaning that the probe state is placed right on the top of the opening. Equation~(\ref{Dysldos1c}) further simplifies to~\footnote{this is consistent with the expression derived in~\cite{SokZel89} for the $T$ matrix.}
\begin{equation}
S(E,x) = -\mathrm{Im} \frac{\langle x|G_0(E)|x\rangle}{1+i\Gamma\langle x|G_0(E)|x\rangle}
\,.
\label{ubersimpS}
\end{equation}
It is clear from the previous expression that here one needs an estimate of the full Green's function of the closed system, which may be obtained in different ways, depending on whether the spectral statistics is random, or it exhibits localization. 
\begin{figure}[tbh!]
\centerline{
\includegraphics[width=7cm]{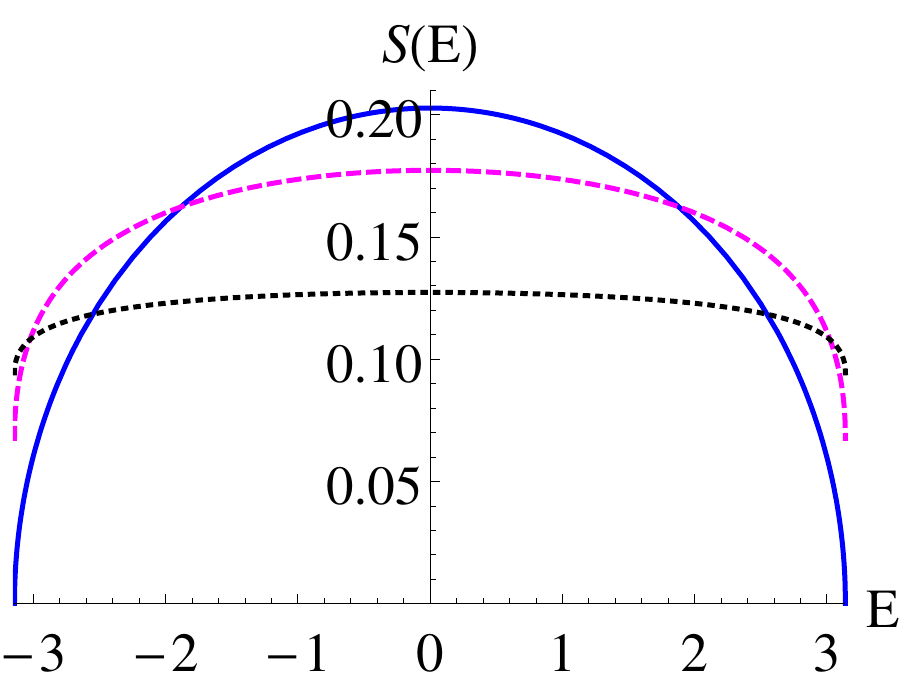}
}      
\caption{Prediction~(\ref{ubersimpS}) for the local density of states for a GOE system, where 
$G_0(E)$ is approximated with the ensemble average $\overline{G}_0(E)$. 
 Solid line: closed system (semicircle rule). Dashed line: open system, $\Gamma=0.5$. Dotted line: open system, $\Gamma=1$.}
\label{onecpred}
\end{figure}
If there are no deviations from RMT statistics,
Eq.~(\ref{ubersimpS}) can be averaged over the RMT ensemble corresponding to the appropriate 
symmetry class:  for example, 
the Green's function of the closed system may be approximated with the average 
of its trace~\cite{SokZel89}
$\overline{G}_0(E)  =
\frac{2}{b^2}\left[E-\sqrt{E^2-b^2}\right]$ for the Gaussian Orthogonal Ensemble (GOE, here $b$ is the range of energies in exam), whose imaginary part is the well-known semicircle rule for the envelope of the 
density of states.
 At this point, one can use a mean-field approximation to  obtain the averaged Eq.~(\ref{ubersimpS}) by just replacing $G_0(E)$ with $\overline{G}_0(E)$.
As a result,
the properly normalized local density of states $\frac{S(E)}{\int_{\Delta E}S(E)dE}$ tends to flatten and becomes increasingly close to a uniform distribution with the coupling to the opening (Fig.~\ref{onecpred}). Since the local spectrum of a RMT-type closed system is meant to have no energy dependence in the first place, we can deem the effect of the opening not dramatic in this case, and withhold further investigation.  

We are faced with a more interesting scenario if the system exhibits scarring: the prediction for the local density of states is still given by Eq.~(\ref{ubersimpS}),
but this time $G_0(E)$ is expected to deviate from RMT ensemble statistics, and to depend on the energy. Thus, rather than taking averages, a semiclassical estimate is attempted for the envelope of the local Green's function at the scar. 
Assume a discrete, non-degenerate spectrum for $H_0$, and 
recall that the autocorrelation function of the closed system is defined as
\begin{equation}
 {\cal{A}}(t)=  \langle x|U_0^t|x\rangle
= \sum_n|\langle x|\psi_n\rangle|^2e^{-iE_nt} 
\,.
\label{autocorrfct}
\end{equation} 
The goal is now to express $G_0(E)$ and  thus $G(E)$ in terms of  ${\cal{A}}(t)$,
for which a 
semiclassical approximation ${\cal{A}}_p(t)$ near the periodic orbit $p$ 
has been available for many years~\cite{KapHel98,Kap99,Cr_Lee}. This is based on 
 the propagation of a wavepacket $|x\rangle$  under the linearized dynamics, and therefore from the contraction/stretching of $|x\rangle$ by
the stable/unstable manifolds (that form an angle of cotangent $Q$), resulting in an overall increase of the variance 
of $|x\rangle$ that depends on the stability exponent $\lambda_p$ \cite{Kap99,Cr_Lee}:
\begin{equation}
{\cal{A}}_p(t) = \frac{e^{-i\phi t}}{\sqrt{\cosh\lambda_p t+iQ\sinh\lambda_p t}}
\,.
\label{CatAp}
\end{equation}
Using the integral representation of the principal value~\cite{GreReinQED} and the above definition of ${\cal{A}}(t)$,
the expectation value of the Green's function $G_0(E)$ may be written as
\begin{eqnarray}
\nonumber
\langle x|G_0(E)|x\rangle &=& \sum_n|\langle x|\psi_n\rangle|^2 P\frac{1}{E-E_n} 
-i\pi\sum_n|\langle x|\psi_n\rangle|^2 \delta(E-E_n)  \\ \nonumber
&=&  \sum_n|\langle x|\psi_n\rangle|^2\frac{i}{2}\int_{-\infty}^{\infty} dt\,\mathrm{sgn}(t)e^{-i(E-E_n)t}
- \sum_n|\langle x|\psi_n\rangle|^2\frac{i}{2}\int_{-\infty}^{\infty}dt\,e^{-i(E-E_n)t} 
\\ 
&=&  -\frac{i}{2}\int_{-\infty}^{\infty}dt{\cal{A}}(t)\mathrm{sgn}(t)e^{iEt} 
- \frac{i}{2}\int_{-\infty}^{\infty}dt{\cal{A}}(t)e^{iEt}   \equiv  R_0(E) - iS_0(E)
\, .
\label{reacldos}
\end{eqnarray}
Since ${\cal{A}}(-t)={\cal{A}}^*(t)$ [and ${\cal{A}}_p(-t)={\cal{A}}^*_p(t)$], 
the two terms of Eq.~(\ref{reacldos})
are respectively real and pure imaginary. While the latter is identified with the local density of states, the former [$R_0(E)$] is usually referred to as reactance in the electromagnetism/microwave cavity literature~\cite{ZhengOtt06,AnlageImp05}. $S_0$ and $R_0$ are thus plugged into Eq.~(\ref{ubersimpS}) for the local densiy of states of the open system.
The advantage of this approach is that 
the semiclassical autocorrelation function ${\cal{A}}_p(t)$ is finally all we need to approximate the local Green's function.\footnote{Equation~(\ref{CatAp}) is derived for the fixed point of a map. Therefore, time is discrete, and the integrals in the last line of Eq.~(\ref{reacldos}) are an approximation, when ${\cal{A}}(t)$ is replaced with  ${\cal{A}}_p(t)\,.$}
\section{Numerics}
The above predictions are now numerically tested on the perturbed cat map~\cite{creagh}
$(q',p')=(q+p-\epsilon\sin 2\pi p, q+2p)\, \mathrm{mod}\,1$,
 whose classical dynamics is fully chaotic on the unit torus, and has an unstable fixed point at the origin
 that gives rise to a scar in the quantization. The unitary quantum map $U_\epsilon=U_0V_\epsilon$ is the product of two $N\times N$ matrices
 respectively describing the linear propagation [of entries $\left<q_j|U_0|q_k\right>~=~N^{-1/2}e^{i\pi/4}e^{2\pi Ni(q_j^2-q_jq_k+q_k^2/2)}$], 
 and a nonlinear kick 
 [$\left<q_j|V_\epsilon|q_k\right>=\sum_{p_m}\frac{1}{N}e^{Ni\left(-\epsilon\cos2\pi p_m+2\pi(q_j-q_k)p_m\right)}$], where the dimension $N$ of the Hilbert space is 
related to the effective Planck constant as $h=1/N$.     
The state $|a\rangle$ identifying the opening is a minimum-uncertainty Gaussian wavepacket 
centered at the periodic orbit:
\begin{equation}
\langle q|a\rangle = \left(\frac{1}{\pi\hbar^2}\right)^{1/4}e^{-(q-q_0)^2/2\hbar+ip_0(q-q_0)/\hbar}
\,,
\label{wpack}
\end{equation}

so that a
non-unitary quantum propagator is realized as~\footnote{The subunitary part of the propagator is implemented as a power series, which is truncated when sufficient convergence of the local Green's function~(\ref{cpudos}) has been achieved.}   
\begin{equation}
U =  e^{-\Gamma\left|a\right>\left<a\right|}U_\epsilon
\,.
\label{eq:incl_leaks}  
\end{equation}  
A sample of 2310 eigenvalues $\varepsilon_n-i\gamma_n$
 and left ($\langle\Phi_n|$) and right ($|\Psi_n\rangle$) eigenstates is produced by numerical diagonalization of 11 realizations of the 
 matrix $U$, where $N$ is varied from 200 to 220 (even numbers). The kick strength is set to
 $\epsilon=0.1$\,.
Local density of states and reactance are then computed 
as real and imaginary parts of the local Green's function
\begin{equation}
 \langle x|G(E)|x\rangle~=~\sum_n h_n(x)\frac{\gamma_n+i(E-\varepsilon_n)}{\gamma_n^2+(E-\varepsilon_n)^2} 
\,,
\label{cpudos}
\end{equation}
with $h_n(x)~=~\frac{\langle x|\Psi_n\rangle\langle\Phi_n|x\rangle}{\langle\Phi_n|\Psi_n\rangle}$, where $|x\rangle=|a\rangle$\,. 

\begin{figure}[tbh!]
\centerline{
(a)
\includegraphics[width=3.8cm]{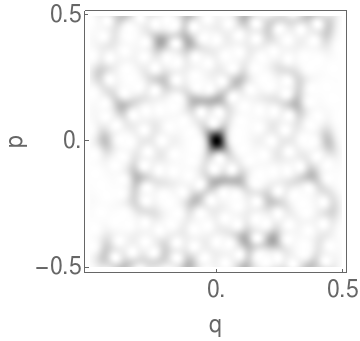}
\hspace{0.05cm}
(b)
\includegraphics[width=3.8cm]{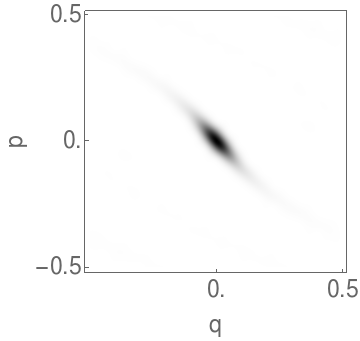}
}
\centerline{
(c)
\includegraphics[width=8.5cm]{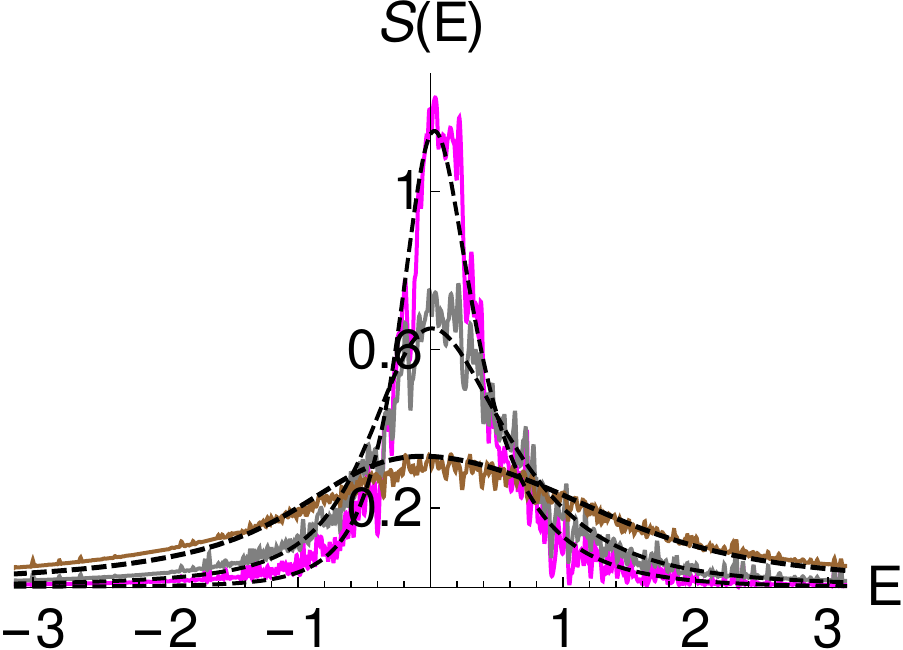}
}
\centerline{
(d)
\includegraphics[width=8.5cm]{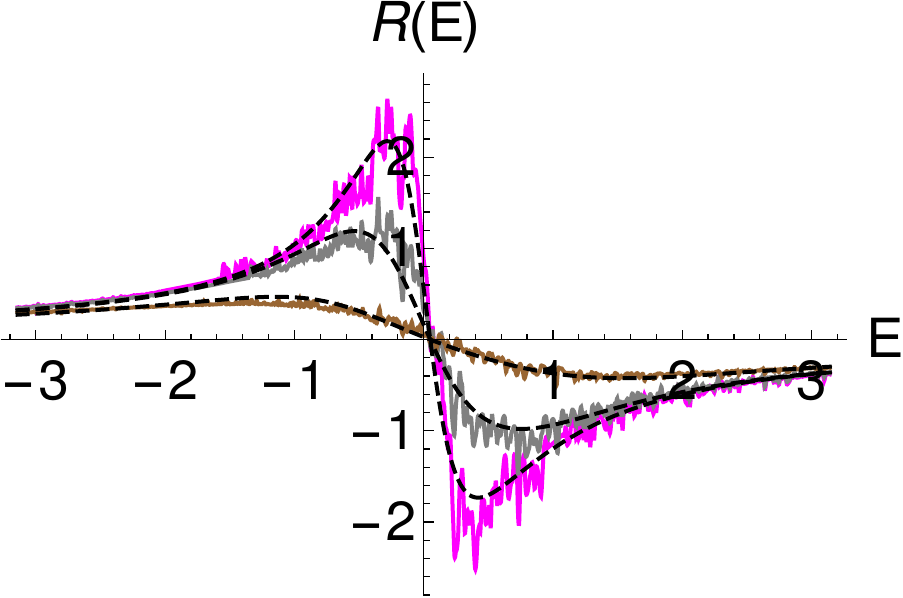}
}
 \caption{(a) The Husimi projection of a scarred eigenfunction of the closed quantum cat map; (b) 
 The projection of the corresponding right eigenstate
 for the open map~(\ref{eq:incl_leaks}).  
 (c) Local density of states of the same system with $\Gamma=0.05$ (highest peak), $\Gamma=0.07$ (middle peak), $\Gamma=0.15$ (lowest peak); (d) Reactance for the same simulation, with the coupling parameters in the same order from the highest peak/ deepest trough to the lowest peak/shallowest trough.
 Solid lines: numerical results; Dashed lines: theoretical estimates.}       
\label{catopnldos}
\end{figure}
Figure~\ref{catopnldos}(c)-(d) shows that the numerically evaluated $S(E)$ and $R(E)$ closely follow their semiclassical expectations,
obtained as real and imaginary parts of Eq.~(\ref{ubersimpS}), with the local Green's function of the closed system being given by  Eq.~(\ref{reacldos}),
and ${\cal{A}}(t)$ is replaced by ${\cal{A}}_p(t)$, as in Eq.~(\ref{CatAp}).
Weak to moderate coupling to the opening 
gradually lowers the peak of the normalized local density of states, and the agreement between theory and numerics is quantitative with no fitting parameters. The numerically computed reactance, that has no normalization, matches the semiclassical estimate up to a multiplicative constant, and it is also 
increasingly smoothened by the opening.     
For large absorption, 
one can indeed observe the predicted flattening of the peak, as the envelope of the local density of states gradually returns to the RMT-predicted uniform distribution. Remarkably, the asymmetry of the envelope, originally due to the non-orthogonality of the stable/unstable manifolds, also tends to disappear with a strong enough coupling to the opening.   
In this regime the sole parameter $\Gamma$ needs to be rescaled 
in order for the theoretical prediction to fit the numerically obtained local density of states. This is ascribed to the quantization of the subunitary evolution operator~(\ref{eq:incl_leaks}), that coincides with the propagator of the non-Hermitian Hamiltonian~(\ref{multcH}) only for weak coupling.        

The progressive shift from an energy-dependent, sharply peaked local density of states to a uniform distribution is due to the losses that cause the spectral linewidths of the scarred eigenstates to widen significantly at large couplings with the opening. A similar trend has been observed experimentally with the distribution of phases of the scattering matrix ($S$) in a microwave cavity with absorption~\cite{Kuhl05}. In appearance, though, this behavior clashes with what observed from the Husimi distributions of the scarred eigenstates, where, instead, localization seems to be enhanced by the opening [Fig.~\ref{catopnldos}(a)-(b)]. 
This phenomenon has been extensively investigated and interpreted using a mode-mode interaction picture: as dissipation increases, the few scarred eigenstates (`doorway states')~\cite{SokZel97,RichSScars} become localized and short-lived, while they separate from a multitude of long-lived eigenstates (`trapped resonances')~\cite{SokZel89,rotlet,ISSO94}.

While a thorough treatment of the multiple-channel problem is deferred to a future publication,
 we want to show that the above behavior of the local density of states is not peculiar of the single-channel opening, but more generic. 
Thus, a numerical computation of the local density of states is performed for
a quantum map that has much in common with simulated optical resonators, given its property of time-reversal symmetry, and its opening according to the Fresnel law of refraction. The map
$(q',p')=(2q+p, 3q+2p)\, \mathrm{mod}\, 1$
 is fully chaotic on the unit torus (details are given in ref.~\cite{han_ber}) and has a fixed point at the origin, which produces a scar in the
quantization $\left<p_j|U_0|p_k\right>=N^{-1/2}e^{i\pi/4}e^{2\pi Ni(p_j^2-p_jp_k+p_k^2)}$\,.
The opening is realized 
in the short-wavelength regime~\cite{SchomKeat,LRLK} assuming transverse magnetic (TM) polarization, as a diagonal matrix $R$ of entries
\begin{equation}
R_{jk} = \delta_{jk}\sqrt{r_{\mathrm{TM}}(p_j)}
\, ,
\label{rmat}
\end{equation}  
where
\begin{equation}  
r_{\mathrm{TM}} = \left(\frac{\sqrt{1-p^2}-\sqrt{n^{-2}-p^2}}{\sqrt{1-p^2}+\sqrt{n^{-2}-p^2}}\right)^2
\,,
\label{TMFresnel}
\end{equation}
for $|p|<1/n$, and $r_{\mathrm{TM}}=1$ otherwise. 
The subunitary propagator is then just given by the matrix product $U=RU_0$.     
\begin{figure}[tbh!]
\begin{center}
\includegraphics[width=8cm]{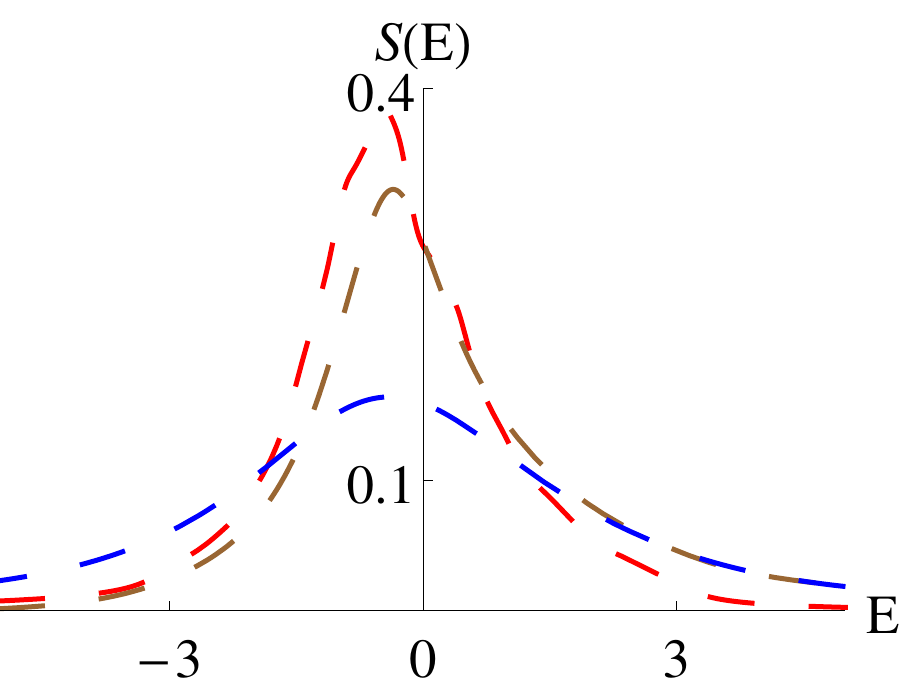}
\end{center}
\caption{Simulation of the quantum map with the Fresnel opening~(\ref{rmat}): the local density of states of this system, numerically evaluatued using the whole computed spectrum, from Eq.~(\ref{cpudos}), with three different refractive indices (from the highest peak down) $n=3.5, n=2.5, n=1.5$.}
\label{opticalmap}        
\end{figure}  
The leaking region now consists of a whole horizontal strip of the phase space, which encloses the fixed point. When the local density of states is evaluated numerically using Eq.~(\ref{cpudos}), taking a coherent state centered at the origin  (of the form~(\ref{wpack}) but projected onto $p-$space) as the probe state $|x\rangle$, the outcome qualitatively resembles what predicted and observed in the case of a single-channel opening [Fig.~\ref{opticalmap}]:
as the open strip of the phase space is widened and the loss is increased with a smaller refractive index, the initially peaked local density of states becomes more
regular, and tends toward a uniform distribution.
\section{Conclusion}
The local density of states of a chaotic system bears information on the spectral statistics, as well as the quantum dynamics in the vicinity of the probe state, since it is directly related to the autocorrelation function. While results and predictions for the envelope of the local spectrum based on both Random Matrix Theory (RMT) and semiclassical treatments have been known for decades in the realm of closed systems, the present work addresses chaotic systems coupled to an opening. Using a self-consistent equation for the Green's function, it has been possible to relate the local spectrum of the open system to the local Green's function of the closed system, and the open channels. That allows us to predict the local density of states, by means of ensemble averaging, if the statistics of the closed Hamiltonian follows RMT, or else by a proper generalization of known semiclassical predictions, if the system deviates from RMT due to scarring.

With the intent of proceeding by steps, we have mainly turned our attention to single-channel openings, where one can clearly see that probe state and open channel must be dynamically correlated, in order for the Green's function and hence the local spectrum to be sensibly modified by the opening. 
In general, significant changes in the spectrum are only appreciable at energy scales of the order of- or greater than the typical linewidth. Spectral envelopes of RMT Hamiltonians 
do not show strong energy dependence either for a closed system or for an open one.    
On the opposite, the distinctively peaked envelope that characterizes scarring is progressively flattened with stronger couplings to the opening, and deviations from RMT are asymptotically suppressed, as the typical spectral linewidth grows toward the order of the energy range.
While this `return to randomness' and fast decorrelations can be mechanistically explained by the increased  linewidths of the scarred eigenstates that result in smaller contributions to the local spectrum, it is less obvious from the Husimi distributions, where scars appear to be enhanced by the losses.      
Numerical realizations of hyperbolic maps on a torus confirm this picture and the theoretical expectations,
for both a single- and a multiple-channel opening inspired by refraction in optical microresonators.






\acknowledgments
The author acknowledges financial support from NSFC (Grant
No. 11750110416-1601190090).

\end{document}